# Orbit–orbit photonics: Harnessing vortex–trajectory interplay for light manipulation

Raghvendra P. Chaudhary[1], Imon Kalyan[1], and Nir Shitrit[1*]

[1]*School of Electrical and Computer Engineering, Ben-Gurion University of the Negev, Be'er Sheva 8410501, Israel*

[*]Corresponding author. E-mail: nshitrit@bgu.ac.il

## Abstract

Light can carry a spin angular momentum, an intrinsic and extrinsic orbital angular momentum, associated with a circular polarization, optical vortex beams, and varying beam trajectories, respectively. The interplay between these momenta yields the spin–orbit interaction of light, in which the spin (circular polarization) controls the spatial (orbital) degrees of freedom of light: either the extrinsic (trajectory) or the intrinsic orbital angular momentum (vortex). While the well-known spin–orbit interaction of light plays a crucial role in nano-optics by providing spin-controlled light manipulation, the interaction between the intrinsic and the extrinsic orbital angular momentum—*the orbit–orbit interaction of light*—has remained elusive. In this interplay, the helical phase fronts of optical vortices control the spatial trajectory of light, giving rise to vortex-dependent shifts of optical beams. We report the orbit–orbit interaction of light in a plasmonic ellipse cavity, whose unique geometry facilitates the interplay when a vortex is considered in one of the foci of the ellipse. In this configuration, the orbit–orbit interaction is achieved by the interplay between the vortex of the source and the ellipse-induced transverse shift of the source beam, positioned at one of the focal points—thus inducing transverse vortex-dependent shifts at the second focal point. Strikingly, the orbit–orbit interaction of light significantly enhances the toolbox available for controlling light by leveraging the *manifold* orbital angular momentum states for vortex-controlled light manipulation—in contrast to light manipulation based on the spin–orbit interaction, which exploits the *binary* polarization helicity.

## 1. Introduction

Light carries both spin and orbital angular momenta—dynamical properties determined by its polarization and spatial degrees of freedom, respectively [1,2]. Optical beams can carry three types of angular momentum: a spin angular momentum, an intrinsic orbital angular momentum (OAM), and an extrinsic OAM [1–3] [Fig. 1(a)]. The spin angular momentum of light originates from the rotation of the electric and magnetic fields in a circularly polarized light beam [4,5], as reasoned by Poynting [4]—more than a century ago. Limited by two polarization states, this intrinsic angular momentum is determined by the handedness of the circular polarization—the polarization helicity $\sigma = \pm 1$, corresponding to right and left circularly polarized light, respectively [1–5]. By contrast, only around 30 years ago, Allen *et al.* [6] discovered that the intrinsic OAM of light is naturally carried by optical vortex beams with phase singularities in their helical phase fronts. The intrinsic OAM, which is associated with the spatial structure of the beam and is relative to its center of gravity, is determined by the topological charge (TC) of the vortex $l = 0, \pm 1, \pm 2, \ldots$—an integer representing the phase increment around the phase singularity [6–12]. Analogous to ubiquitous vortices in nature such as hurricanes and whirlpools, vortex beams are screwed along the propagating direction, resulting in a dark core at the beam center. Optical beams can also carry an extrinsic OAM, which appears in optical beams propagating at a varying distance from the coordinate origin and is related to the motion of the center of gravity of the beam [13]; it is analogous to the mechanical angular momentum of a classical particle, given by the cross-product of the transverse position of the beam center and its wave vector [Fig. 1(a)], which characterize the beam trajectory [3,13]. These three separable observable components of the optical angular momenta are expressed by values per photon in $\hbar$ units, where $\hbar$ is the reduced Planck's constant [1,2].

The interplay and mutual conversion between these three types of optical angular momentum yield the spin–orbit interaction (SOI) of light—a striking optical phenomenon in which the spin (circular polarization) controls the spatial (orbital) degrees of freedom of light [3,14–16]. The SOI includes the interaction between the spin angular momentum and the extrinsic OAM (trajectory) [Fig. 1(a)], which results in a helicity-dependent position or momentum of light, known as the spin Hall effect of light [3,17–23]. Moreover, the coupling between the spin angular momentum and the intrinsic OAM (vortex) [Fig. 1(a)] yields a spin-to-OAM conversion, i.e., helicity-dependent optical vortices [3,21,24–34]. Remarkably, the SOI of light plays a crucial role in the new reality of nano-optics by providing a toolbox for spin-controlled light manipulation [3]. However, while the SOI of light is well known and has been widely studied, the interaction between the intrinsic OAM and the extrinsic OAM—the orbit–orbit interaction (OOI) of light [see Fig. 1(a)]—has remained elusive; moreover, previous studies [35–40] focused on free-space optics, such that nanophotonic systems and their richness remained unexplored. In this nontrivial interplay, the helical phase fronts of optical vortices control the spatial trajectory of light, giving rise to vortex-dependent shifts (VDSs) of optical beams. Strikingly, the OOI of light significantly enhances the toolbox available for controlling light by leveraging the *manifold* OAM states ($l = 0, \pm 1, \pm 2, \ldots$) for vortex-controlled light manipulation, in contrast to SOI-based light manipulation [3], which exploits the *binary* polarization helicity ($\sigma = \pm 1$).

Here, we report the demonstration of the OOI of light in a plasmonic ellipse cavity (PEC), providing a nanophotonic realization of this effect, whose unique geometry facilitates the OOI when a vortex is considered in one of the foci of the ellipse. In this configuration, the OOI between the intrinsic OAM and the extrinsic OAM is achieved by the interplay between the vortex of the source and the ellipse-induced transverse shift of the source beam, positioned at one of the focal points. The OOI of light in the PEC induces transverse VDSs, i.e., shifts that depend on both the vortex helicity and strength, at the second focal point [Fig. 1(b)]. Moreover, we demonstrate a proof-of-concept for all-optical information processing via on-chip OAM demultiplexing based on the OOI by encoding information via different vortex states in the superposed source beam and decoding it via the spatially separated VDSs [Fig. 3(b)]. By leveraging the manifold OAM states, the OOI of light opens a new paradigm for vortex-controlled light manipulation; moreover, the OOI offers great potential for OAM-supported applications, including high-bandwidth optical communications and quantum information processing, enhanced resolution in imaging and microscopy, control of matter by optical trapping and tweezing, and many more.

## 2. Results: Numerical simulations and a theoretical model

To demonstrate the concept of the OOI of light, we investigated a physical system of a two-dimensional plasmonic cavity, defined by an ellipse Bragg grating, in which a source of intrinsic OAM is considered in one of the foci of the ellipse [Fig. 1(b)]. The ellipse cavity was chosen due to its unique geometric property, i.e., a curve on a plane surrounding two focal points, such that for every point on the curve, the sum of the distances to the two focal points is constant. Therefore, if we place a source without an intrinsic OAM—a Gaussian beam—at one of the foci of the ellipse cavity, then, according to Fermat's principle of least time, we expect the source to be perfectly 'imaged' to the second focal point [Fig. 1(b)]. Conversely, if the source carries an intrinsic OAM, we expect to reveal transverse VDSs at the second focal point [Fig. 1(b)]. This observation is the manifestation of the OOI of light between the intrinsic OAM and the extrinsic OAM, where the helical phase fronts of optical vortices control the spatial trajectory of light. Uniquely, the geometric property of the ellipse makes it an excellent candidate to study the OOI via the interplay between the vortex of the source and the ellipse-induced transverse shift of the source beam, positioned at one of the focal points. In the OOI of light, the sign of the shift depends on the vortex helicity and its magnitude depends on the vortex strength, such that higher values of intrinsic OAM result in larger shifts. Note that we chose a plasmonic system that supports propagating surface-confined waves—surface plasmon polaritons (SPPs)—as it offers a simplified realization by reducing the dimensionality of the system.

We performed numerical simulations to observe the VDSs in the PEC. We used a finite-difference time-domain algorithm (Lumerical) to calculate the electromagnetic near fields of the PEC. We considered an ellipse with semi-major and semi-minor axes of $a = 10$ μm and $b = 8$ μm, respectively (the center-to-focus distance is $\sqrt{a^2 - b^2} = 6$ μm) [Fig. 1(c)]. To define the PEC, we introduced a dielectric Bragg grating at the metal–air interface [Fig. 1(c)]—a three-layer configuration in which the effective mode index of SPPs can be robustly engineered via the height of the dielectric [polymethyl methacrylate (PMMA)] cladding layer [41,42]. We considered an air–PMMA–gold structure to realize the ellipse Bragg grating, that defines the PEC by encircling it and provides high reflectivity for the SPPs by satisfying the Bragg condition. We considered a 130-nm-thick PMMA grating with a period of 384 nm and a duty cycle of 50% (linewidth of 192 nm), on top of a 50 nm gold film [Fig. 1(c)], at the wavelength of 810 nm (the period of the ellipse Bragg grating is approximately half the SPP wavelength). For the source, we realized an optical vortex by introducing Laguerre–Gaussian beams—cylindrically symmetric solutions to the wave equation carrying an OAM of $l\hbar$ per photon [1,2,6,7]. We considered normally incident beams with a beam waist (radius) of 1.4 μm and a varying TC $l$ (Fig. 2 insets). This vortex source was positioned at one of the foci of the ellipse—without loss of generality, we chose the left focal point—and the incident polarization was set to be linear ($y$ polarization with zero spin angular momentum). To enhance the coupling efficiency of the incident free-space light to SPPs for the multiple vortex modes, a scatterer was introduced at the left focal point of the ellipse cavity [Fig. 1(c)]—the position of the source beam. The top-illuminated hybrid scatterer (130 nm PMMA thickness) is composed of an inner nanodisk (diameter of 794 nm—the SPP wavelength) and outer concentric rings (3 rings with a period of 794 nm and a duty cycle of 50%). While the inner nanodisk strongly overlaps with the low-order vortices, the outer rings significantly overlap with the high-order vortices (the profiles of the vortex modes are shown in Fig. 2 insets). All structure parameters were optimized to maximize the intensities of the VDSs.

Figure 2 shows the simulated results of the near-field intensity distributions of the PEC for different TCs $l$ of the source. The test case of a source without a vortex, i.e., a Gaussian beam with $l = 0$, shows the expected result—the source is 'imaged' to the right focal point due to the unique geometric property of the ellipse [Fig. 2(a)]. Conversely, once the vortex is introduced by the source beam, VDSs from the right focal point are clearly observed [Figs. 2(b)–2(k)]. This observation stems from the OOI of light between the intrinsic OAM (vortex) and the extrinsic OAM (trajectory) in the PEC, which enables the interplay between the vortex of the source beam and its transverse shift, induced by the ellipse, once the source is positioned at one of the focal points. The observed VDSs can be referred to as the orbital Hall effect of light [35–40,43], wherein the helical phase fronts of optical vortices control the spatial trajectory of light, thus paving the way to vortex-enabled light manipulation. Figure 3(a) shows the dependence of the transverse shift, extracted from the intensity patterns, on the TC of the source beam (we track the spot corresponding to zero transverse shift for $l = 0$ and maximum intensity near the right focal point). In this effect, the sign of the shift depends on the vortex helicity, similar to the spin Hall effect of light, wherein the sign of the shift depends on the polarization helicity [3]; the VDSs corresponding to $l = \pm 1$ are equivalent to the helicity-dependent spin Hall shifts (corresponding to $\sigma = \pm 1$). However, here, the shift also depends on the vortex strength ($|l| = 0,1,2,\ldots$), such that higher values of OAM result in larger shifts. Since the spin Hall shifts are exceedingly small and subwavelength, their experimental observation requires nonstandard measurement methods (accumulation of the effect through many multiple reflections [20], ultrasensitive quantum weak measurements [19], etc.); strikingly, the reported VDSs [Fig. 3(a)] provide a route to overcome this arduous task by utilizing the high-order vortex states. We also show that the longitudinal shift is independent of the TC of the source beam and is effectively zero [Fig. 3(a) inset]; these results reveal the purely transverse nature of the VDSs in the OOI of light (for the smallest TC of $l = \pm 1$, which exhibits the smallest nonzero transverse shift, the ratio between the transverse and longitudinal shifts is around 135). Moreover, as the spin Hall shifts are manifested by $\sigma/k_{\mathrm{SPP}}$ [28,30], where $k_{\mathrm{SPP}}$ is the wave number of SPPs, the VDSs were fitted linearly [Fig. 3(a)], formulated as $-\alpha l/k_{\mathrm{SPP}}$; here, the proportion coefficient $\alpha(a/b)$ represents the geometry of the ellipse, where $a/b$ is its aspect ratio.

To further study the OOI of light, we derived an analytical scalar model to calculate the optical field inside the PEC, so as to quantify the VDSs—the observable quantity of the vortex–trajectory interplay. We considered an ellipse curve, centered at the origin, which is defined by the equation ${x_0}^2/a^2 + {y_0}^2/b^2 = 1$, where $(x_0,y_0)$ are the coordinates on the curve and $a$ and $b$ are the semi-major and semi-minor axes, respectively [Fig. 5(a)]. We considered a point source positioned at one of the foci of the ellipse (without loss of generality, we chose the left focal point), where the plasmonic source carries an intrinsic OAM, i.e., a surface-confined propagating vortex beam with a helical phase front of $\exp(il\varphi)$ ($\varphi$ is the azimuthal angle with respect to the position of the source). The plasmonic vortex launched from the source propagates along the two-dimensional surface until it meets the ellipse. Within this scalar framework, polarization effects and vectorial field components are neglected, and the SPP is treated as a scalar wave described by a complex field. Material losses are incorporated through the complex SPP wave number. Therefore, the source induces a field on the ellipse curve of

$$E_0(x_0,y_0) = \exp(il\varphi_0)\exp(ik_{\mathrm{SPP}}r_0); \tag{1}$$

here, $r_0(x_0,y_0)$ is the distance between the source and the local point on the ellipse, $\varphi_0(x_0,y_0)$ is the corresponding azimuthal angle [see Fig. 5(a)], and $k_{\mathrm{SPP}}$ is the SPP wave number. While the first term is associated with the helical phase front of the source, the second term is associated with a propagation-induced phase. To mimic a two-dimensional plasmonic cavity defined by the inner region of the ellipse, we considered each point on the ellipse as a secondary source that reflects the light from the original source, thus spatially limiting the light to within the ellipse. This treatment assumes a uniform reflection amplitude along the ellipse boundary and neglects any position-dependent phase shifts arising from the reflection process, thereby providing an idealized cavity response. Therefore, the field generated by all the secondary sources on the ellipse is

$$E(x,y) = \sum E_0(x_0,y_0)\exp(ik_{\text{SPP}}r), \tag{2}$$

where $r(x,y)$ is the distance between a local point on the ellipse and the point of calculation $(x,y)$ [see Fig. 5(a)]; note that the sum represents a coherent summation (i.e., interference) of all secondary sources. In practice, the continuous ellipse boundary is discretized into a sufficiently dense set of points (secondary sources) to ensure numerical convergence of the summation and accurate reconstruction of the resulting interference pattern. This theoretically derived wave interference model can be regarded as the Huygens–Fresnel principle [44]—a construction of a wavefront by secondary wavelets induced by the primary wavefront—with OAM.

We considered an ellipse cavity with semi-major and semi-minor axes of $a = 10$ μm and $b = 8$ μm, respectively (resulting in a center-to-focus distance of $f = 6$ μm). We demonstrated the OOI of light in near-infrared plasmonics, where the propagation length of SPPs at a gold–air interface, at the wavelength of 810 nm is $\sim$40 μm [45–48]; note that the sum of propagation distances from the source (one focal point) to the ellipse, and from the ellipse to the second focal point is $2a = 20$ μm. Based on the derived model, we calculated the intensity inside the PEC (Fig. 4) and extracted the VDSs [Fig. 3(a)], which manifest the OOI of light. The calculated shifts exhibit good agreement with the simulated results [Fig. 3(a)]. These theoretical and numerical results are proof-of-concept demonstrations of the OOI of light—the interplay between the intrinsic OAM (vortex) and the extrinsic OAM (trajectory), represented by the vortex of the source and the inherent ellipse-induced transverse shift of the source beam, positioned at one of the focal points, respectively—in plasmonics. More generally, the OOI-facilitated influence of helical phase fronts of optical vortices on the spatial trajectory of light paves the way for vortex-enabled light manipulation.

This generic toy model also reveals the necessity of the two different types of OAM in the OOI of light. When the source does not carry an intrinsic OAM (i.e., $l = 0$), the source is perfectly 'imaged' to the second focal point, such that the observable quantity of the OOI—the VDS—is zero [see Fig. 5(b)]. On the other hand, when a circle (i.e., $a = b$) is considered instead of an ellipse [Fig. 5(c)], the center-to-focus distance is zero, so the two focal points become indistinguishable. Therefore, the transverse shift of the source is zero, which eliminates the contribution to the OOI from the extrinsic OAM (trajectory), appearing in optical beams propagating at a distance from the coordinate origin. In this case, the model predicts vortex-dependent intensity profiles inside the cavity that exhibit circularly symmetric patterns of rings [Fig. 5(c)]. In analogy to a center-of-mass position, the observed dark spots in the center of the cavity reveal that the VDSs are zero [see Fig. 5(c)].

By virtue of the unlimited OAM states, in principle, optical vortices provide one of the promising solutions for enhancing the capacity of data multiplexing to meet the unprecedented growth in big data. The OAM of light can dramatically boost the channels for data transmission and storage in optical communications [49] and quantum information processing [50], which are considered to be the next generation of high-capacity photonics [51,52]. In quantum optics, spatially shaped modes carrying OAM (e.g., Laguerre–Gaussian modes) represent a high-dimensional orthogonal basis of quantum states of light, assessed with respect to the entanglement of OAM [50], with its tremendous potential for data encryption and security [8,10]; by accessing high-dimensional quantum states via laboratory realizations of OAM using single photons, OAM leverages quantum information processing. As such, optical vortices have the potential to tremendously increase the capacity of communication systems, either by encoding information as the OAM states of the beam or by using OAM beams as information carriers for multiplexing [51–56]. Owing to the inherent orthogonality of OAM beams and the linear nature of the OOI, the vortex–trajectory interplay can be harnessed to enable both the multiplexing and demultiplexing of OAM beams to increase the capacity of optical communication systems. To demonstrate this concept, we study the OOI in the PEC with a source beam comprising a superposition of vortex beams with various TCs. In such a system, the VDSs resulting from the OOI enable demultiplexing by spatially separating each vortex state of the input [Fig. 1(b)]. To extend the on-chip OAM demultiplexing facilitated by the OOI to information processing, we encode information via the different OAM states in the superposed source beam and decode it via the spatially separated VDSs [Fig. 3(b)]. We consider the encoded information as a binary string of 1s and 0s (e.g., an ASCII binary protocol), corresponding to the on and off states, respectively, of the different OAM states used as the basis [56]; without loss of generality, we consider a set of eight orthogonal OAM states ($l = \pm 1, \pm 2, \pm 3$, and $\pm 4$) as the basis for encoding the information using the ASCII binary protocol [Fig. 3(b)]. In this protocol, each OAM state in the chosen basis represents 1 bit of information in the 1 byte (8 bits) of information, representing an alphabet letter or a symbol. The superposed source beam is generated by multiplexing the OAM on states. Owing to the known dependence of the shift on the TC [see Fig. 1(b)], the OOI-based OAM demultiplexing allows us to decode the information;

we restore the binary string by assigning either 1 or 0 to the OAM states in the basis based on whether a shift was observed (1) or not (0) [Fig. 3(b)]. Traditional methods for decoding the information include the process of inner product between the multiplexed signal and the OAM basis; however, here, the information decoding is direct and therefore preferable, as it is inherent in the OOI-based OAM demultiplexing. Therefore, the on-chip OAM multiplexing and demultiplexing enabled by the OOI in the PEC, and the accompanied information processing capabilities, open a new paradigm for integrated photonic devices utilizing the manifold OAM states in a digitized manner.

## 3. Conclusions

In summary, we report the OOI of light—the interaction between the intrinsic OAM (vortex) and the extrinsic OAM (trajectory)—in a PEC, whose unique geometry facilitates the vortex–trajectory interplay when a vortex is considered in one of the foci of the ellipse. In contrast to SOI-based light manipulation [3], which exploits the *binary* polarization helicity, the OOI of light significantly enhances the toolbox available for controlling light by leveraging the *manifold* OAM states for vortex-controlled light manipulation. In this work, we use the term OOI of light in a phenomenological sense, referring to the interplay between the intrinsic and the extrinsic OAM [see Fig. 1(a)], whereby the helical phase fronts of optical vortices influence the propagation trajectory of light; this interpretation is consistent with prior studies on the orbital Hall effect [35–40,43], in which this interplay manifests as vortex-dependent beam shifts. Experiment-wise, the VDSs—the observable quantity of the OOI of light—can be observed using scanning near-field optical microscopy [28,30,34,57,58], leakage radiation microscopy [42,59–61], or fluorescence imaging [42], which are common tools for imaging the propagation of SPPs. Beyond PECs, the OOI can be explored in asymmetric deformed resonators, such as stadium-shaped cavities, or in multifocal geometries, such as polyelliptic cavities, where modified orbit structures or additional focal points may give rise to richer displacement phenomena. While the spin Hall effect of light [17–20] emerged and was studied as the photonic version of the spin Hall effect in electronic systems [62,63], the presented orbital Hall effect of light may leverage and advance analogous studies in electronic (fermionic) systems [64]. Moreover, by utilizing the multiple OAM states as a new degree of freedom in light manipulation, the OOI of light offers great potential for OAM-supported applications; these include high-bandwidth optical communications [49,51–56] and quantum information processing [50], enhanced resolution in microscopy and imaging [65,66], control of matter by optical trapping and tweezing [67–69], enantiomer-selective sensing [70], and many more.

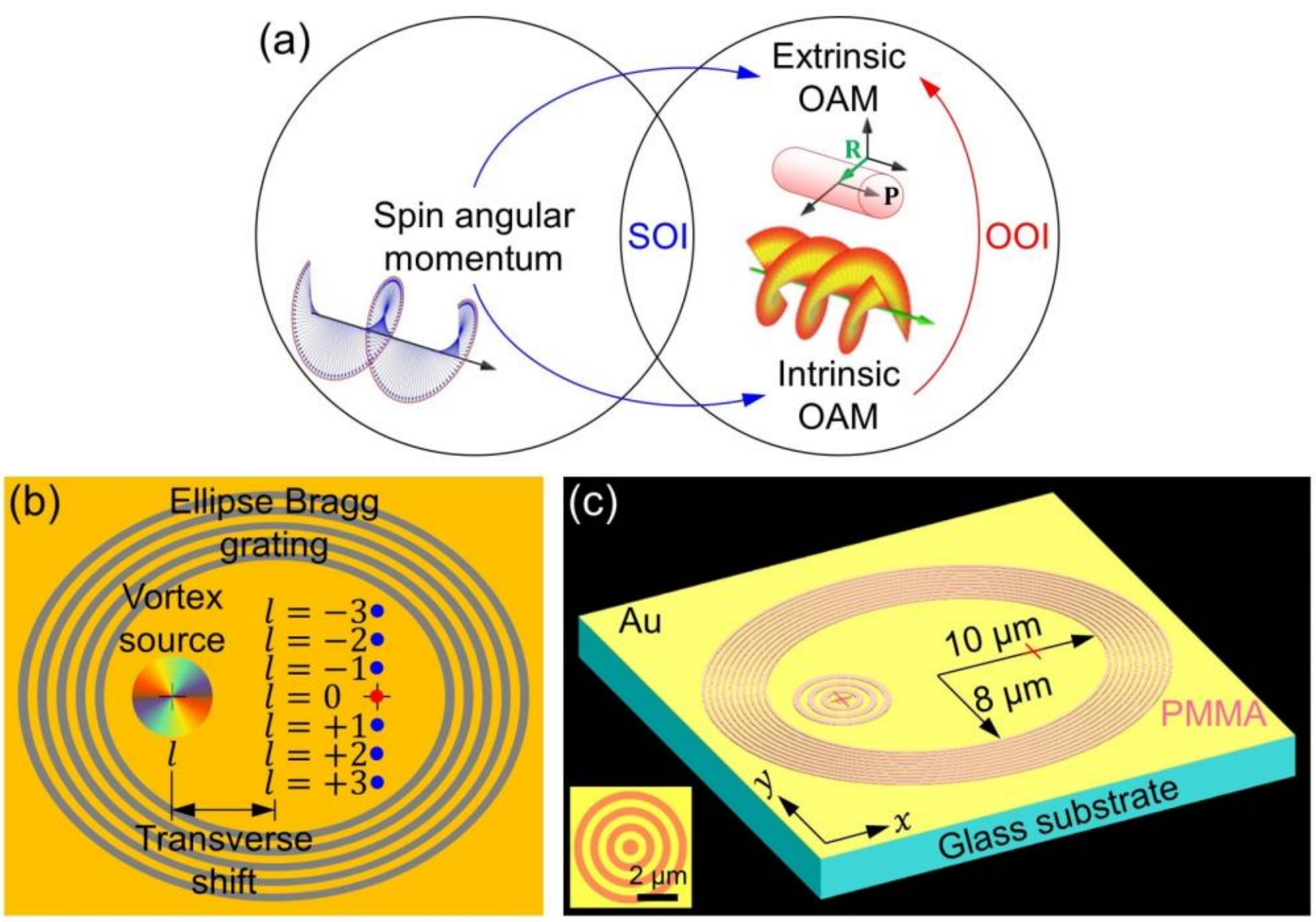

**Fig. 1.** Orbit–orbit interaction of light: Concept and physical system. (a) Light can carry a spin angular momentum (polarization helicity), an intrinsic OAM (vortex), and an extrinsic OAM (trajectory) (**R** indicates the transverse shift of the beam center and **P** is its momentum, such that $\mathbf{R} \times \mathbf{P}$ represents the extrinsic OAM). While the SOI of light regards the interaction between the spin angular momentum and either the extrinsic or intrinsic OAM, the OOI of light arises from the interplay between the intrinsic and extrinsic OAM. The OOI manifests as the influence of helical phase fronts of optical vortices on the trajectory of light and leverages the manifold OAM states for vortex-controlled light manipulation. (b) The OOI of light in a PEC induced by the interplay between the vortex of the source, considered in the left focal point of the ellipse, and the ellipse-induced transverse shift of the source beam. VDSs are the observable quantity of the OOI. (c) Simulated structure. We considered an ellipse with semi-major and semi-minor axes of 10 μm and 8 μm, respectively. We also considered a PMMA Bragg grating, on top of a gold (Au) film. To couple the incident free-space light to SPPs, a top-illuminated PMMA scatterer, composed of an inner nanodisk and outer concentric nanorings (see inset), is introduced at the left focal point of the ellipse cavity.

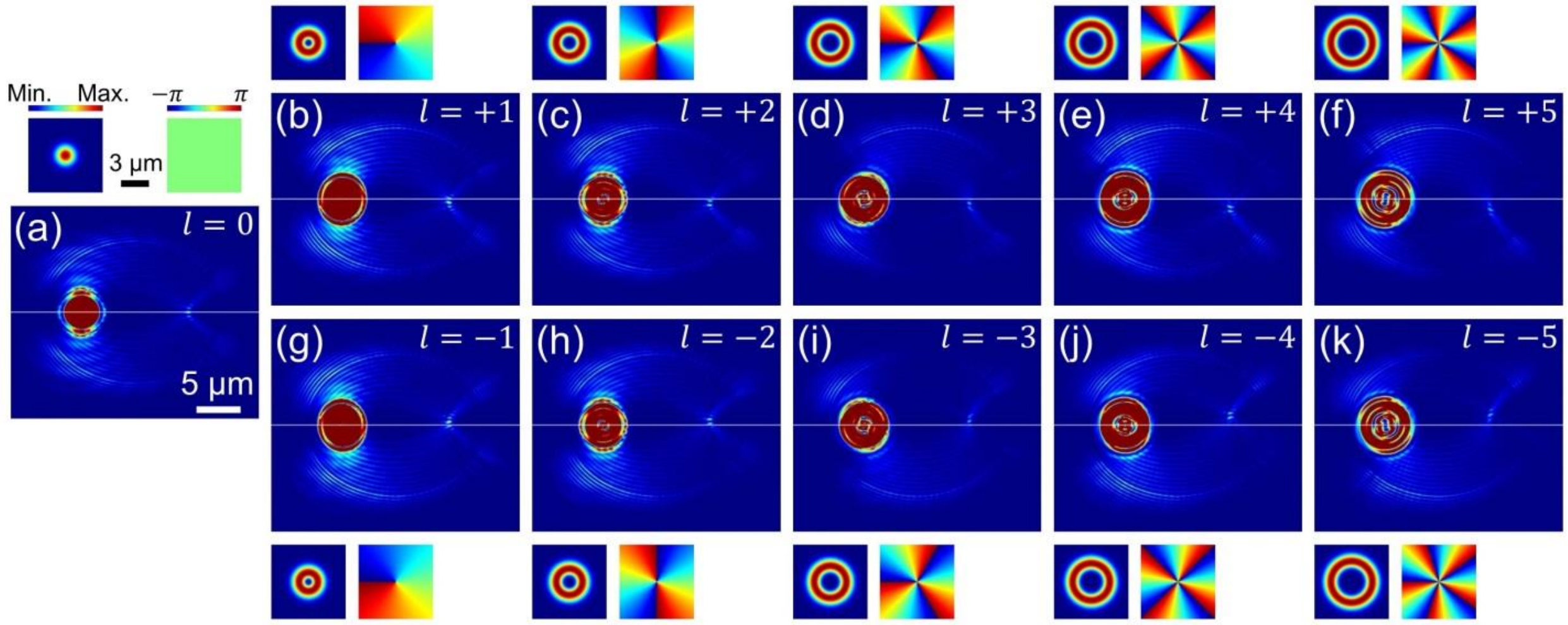

**Fig. 2.** Orbit–orbit interaction of light in a plasmonic ellipse cavity. A vortex source beam with a TC $l$ is positioned at the left focal point of the ellipse, illuminating it at normal incidence at the wavelength of 810 nm. The intensity and phase profiles of the Laguerre–Gaussian source beam are shown in the left and right insets, respectively. (a) Simulated near-field intensity ($|E_y|^2$) map of the PEC for $l = 0$, where a transverse shift from the right focal point is not observed. (b)–(f) Simulated near-field intensity maps for $l = +1, +2, +3, +4,$ and $+5$, respectively. (g)–(k) Simulated near-field intensity maps for $l = -1, -2, -3, -4,$ and $-5$, respectively. The horizontal symmetry lines reveal the VDSs, where positive and negative TCs correspond to negative and positive transverse shifts, respectively. The scale bar in (a) is applicable for all panels.

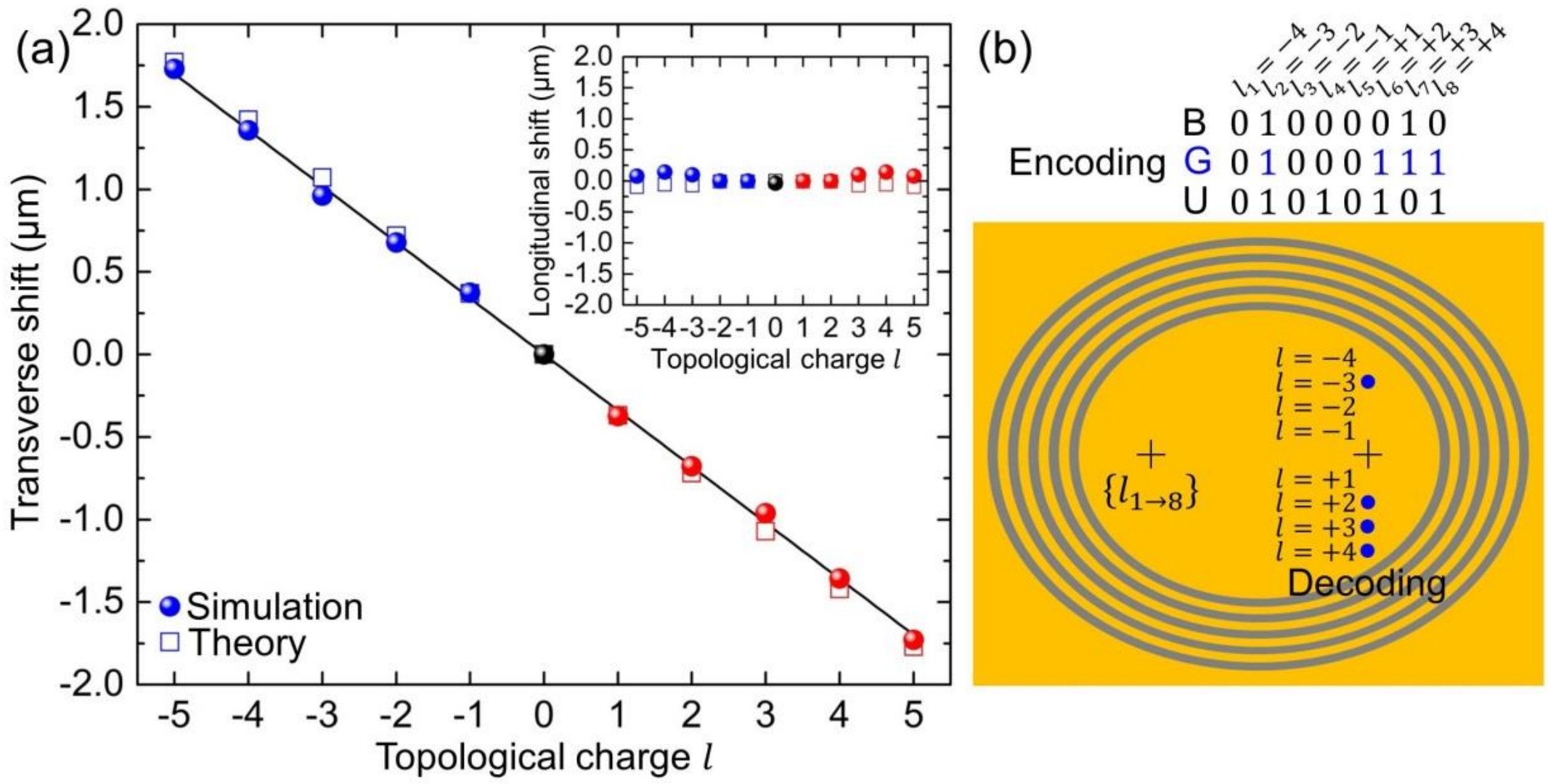

**Fig. 3.** Vortex-dependent shifts in the plasmonic ellipse cavity and information processing based on the orbit–orbit interaction of light. (a) Dependence of the transverse shift on the TC $l$ of the source beam. Dots refer to data from numerical simulations, whereas squares refer to theoretical calculations performed by the wave interference model. The line is the linear fit of the numerical data. The inset shows the dependence of the longitudinal shift on the TC of the source beam. (b) All-optical on-chip information processing based on the OOI of light by encoding information (e.g., using ASCII binary strings) via the different vortex states $\{l_{1\to8}\}$ in the superposed source beam, and decoding the information via the spatially separated VDSs. In this proof-of-concept demonstration, the word 'BGU', represented in the ASCII binary protocol, is sent, with each letter sent separately, by multiplexing the corresponding OAM on states; we then decode the information by observing the VDSs, which directly reveal the ASCII representation of the word via the OAM demultiplexing.

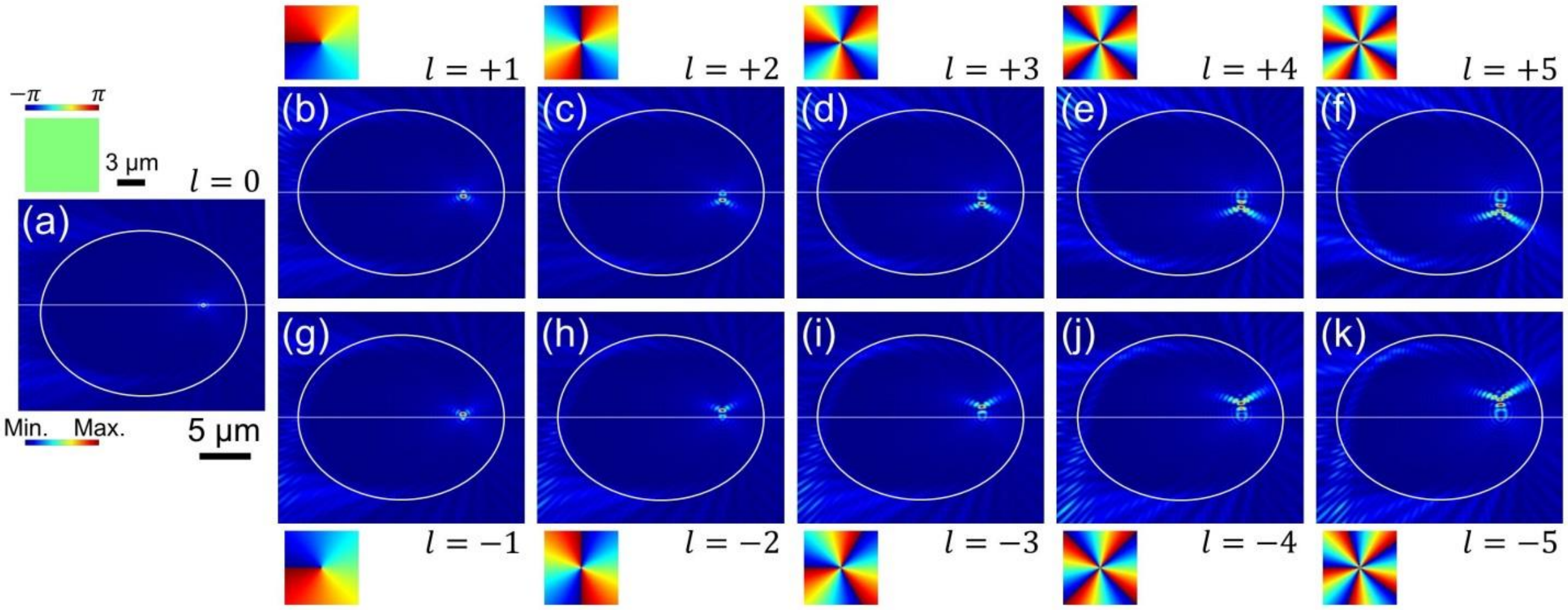

**Fig. 4.** Orbit–orbit interaction of light in a plasmonic ellipse cavity: Theoretical results. A vortex source with a TC $l$ is considered at the left focal point of the ellipse, at the wavelength of 810 nm. The phase profile of the source is shown in the inset. (a) Calculated near-field intensity map of the PEC for $l = 0$, where a transverse shift from the right focal point is not observed. (b)–(f) Calculated near-field intensity maps for $l = +1, +2, +3, +4,$ and $+5$, respectively. (g)–(k) Calculated near-field intensity maps for $l = -1, -2, -3, -4,$ and $-5$, respectively. The horizontal symmetry lines reveal the VDSs, where positive and negative TCs correspond to negative and positive transverse shifts, respectively. The scale bar in (a) is applicable for all panels.

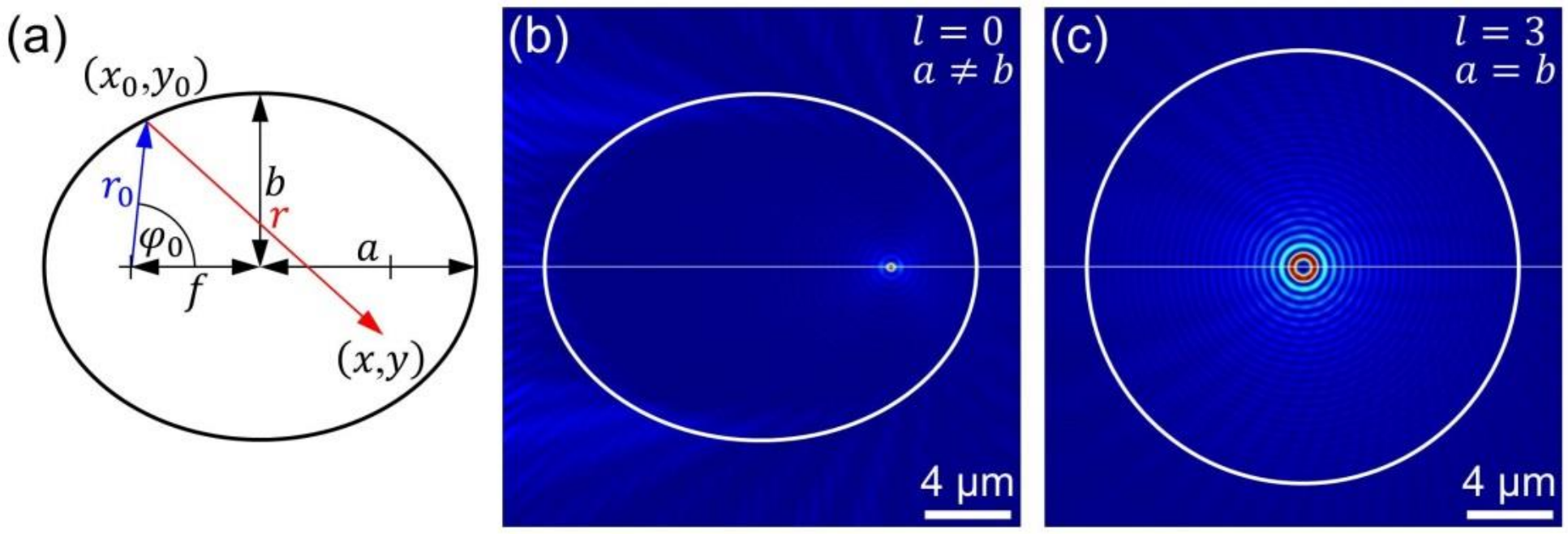

**Fig. 5.** Orbit–orbit interaction of light in a plasmonic ellipse cavity: Necessity of the two orbital angular momentum components. (a) Schematics used to derive the theoretical wave interference model. The vortex source beam is positioned at the left focal point of the ellipse. (b) Near-field intensity map of the PEC for $l = 0$, i.e., without an intrinsic OAM (vortex) of the source beam. (c) Near-field intensity map for $a = b = 10$ μm, i.e., the ellipse transforms to a circle (the calculation was performed for $l = 3$); in this case, where the foci distances are zero, the extrinsic OAM (trajectory) associated with the ellipse-induced transverse shift of the source is eliminated. Both results were calculated by the model. The white lines show the outlines of the ellipse and circle. As indicated by the horizontal symmetry lines, no shifts are observed in these special cases.

**Acknowledgments**

The authors gratefully acknowledge funding from the Israel Science Foundation (ISF) under Grant No. 1785/22.

**Disclosures**

The authors declare no conflicts of interest.

**Data availability**

Data underlying the results presented in this paper are not publicly available at this time but may be obtained from the authors upon reasonable request.